# The Evolution of Cooperation in Business

Dan Ladley, University of Leicester

Ian Wilkinson, University of Sydney and University of Southern Denmark

Louise Young, University of Western Sydney and University of Southern Denmark

## ABSTRACT

The development of cooperative relations within and between firms plays an important role in the successful implementation of business strategy. How to produce such relations is less well understood. We build on work in relational contract theory and the evolution of cooperation to examine the conditions under which group based incentives outperform individual based incentives and how they produce more cooperative behavior. Group interactions are modeled as iterated games in which individuals learn optimal strategies under individual and group based reward mechanisms. The space of possible games is examined and it is found that, when individual and group interests are not aligned, group evaluation and reward systems lead to higher group performance and, counter-intuitively, higher individual performance. Such groups include individuals who, quite differently to free-riders, sacrifice their own performance for the good of the group. We discuss the implications of these results for the design of incentive systems.

**1 Introduction**

The focus of much research and teaching in management concerns decision analysis rather than administration or implementation. As Henry Mintzberg (2005) observes in his book *Managers not MBAs* "MBA students graduate with the impression that management is analysis, specifically the making of systematic decisions and the formulation of deliberate strategies". But management is more; it is about getting things done, about implementing decisions within and across firm boundaries. It involves working with people who need to carry out interrelated activities, using specialized resources and skills in a particular context and who need to communicate, coordinate and cooperate to achieve goals.

This paper considers how management can improve the behavior and performance of groups and teams. Empirical research has identified various characteristics of groups, relations and networks that are associated with improved performance, including the role of cooperative behavior (e.g. de Jong et al 2005, Cummings and Cross 2003, Dyer 2006, Spekman and Davis 2004, Wilkinson 2008). These studies, however, do not tell us how such cooperation can be developed and managed, only the personal and behavioral characteristics that are associated with its presence. More generally, the reasons for cooperative and altruistic behavior, the presence of which has been repeatedly demonstrated in experiments carried out across many cultures (e.g. Henrich et al 2004), is not well understood. Recent developments in biological theories of the evolution of cooperation point to the importance of group based selection and reward mechanisms in driving the emergence of cooperative behavior and superior performance. This paper investigates whether these theories may be applied to the design of business incentive systems that improve cooperation and performance.

To examine the role of group based incentive systems we develop an agent based simulation model of the interactions of individuals within groups. Individuals engage in pair wise iterated games with other group members. A genetic algorithm is used to capture the impact of individual and group based selection and reward systems on behavior. In an exhaustive set of simulations we examine the conditions under which group reward systems are superior to individual based systems in producing better performing, and more cooperative, groups and individuals. Group based systems are demonstrated to be superior, for many, but not all types of games. We argue these results have important implications for the design of intra and inter-firm incentive systems and help explain the results of previous experiments and case studies.

By performing computation experiments we are able to consider a complex setting and a large range of games that would be beyond mathematical tractability (Leombruni and Matteo 2005). Whilst laboratory experiments structured in the same manner as those we carry out, could, in theory, be used to investigate individual specifications. To examine the vast number of empirical experiments represented by the simulations would be practically impossible. Simulation enables us to represent key features of group interaction and incentive systems and to track outcomes over time, under a wide variety of controlled conditions. The simulation results therefore provide a basis for designing future empirical research.

The paper is organized as follows. First, we discuss the role of cooperation in business and review research concerning the evolution of cooperation. We then develop an agent based model of group interaction and evolution and use it to examine the effect of different reward and selection mechanisms. Detailed results are first presented for three illustrative games and then for simulations spanning an exhaustive space of games. Finally, the relevance and implications of our results for the design of intra and inter-firm incentive systems is discussed together with future research opportunities.

## 2 Cooperation in Business

Cooperative relations are increasingly seen as a source of competitive advantage for firms, industries, regions and nations (e.g. Dyer 2006, Spekman and Davis 2004). Within a firm there are many formal and informal groups of people and departments carrying out interdependent activities that co-produce value (Stabell and Fjeldstad 1998). As one manager comments: "We think everything worth doing is done by groups, not by individuals" (Weber et al. 2005, p. 80). The degree of cooperation and coordination within and among groups effects firm performance (Luo et al 2006, Hakansson and Snehota 1995; Rulke and Galaskiewicz 2000; Smith et al, 1995). Similarly the value of cooperation between firms has also been highlighted (Wilkinson 2008; Xu and Beamon 2006; Mattsson 1997; Morgan and Hunt 1994).

Prior research has analyzed the role of cooperation in business and identified various characteristics of individuals, groups, relations and networks associated with superior performance. The means by which groups, relations and networks with these characteristics are produced and maintained is not well understood. Producing such groups is not straight forward because managers and firms cannot directly control or monitor the behavior of all actors involved. Within firms, managers control formal hierarchies and systems, however, informal influence structures also emerge and self-organize. These informal structures evolve as a result of the outcomes and experiences of the various interactions which take place over time. Overall performance is not a linear sum of individual performances, it is co-produced and difficult to trace to, and control through the actions of individuals. Similarly, the performance of firms depends on the actions of others both directly and indirectly and on the types of relations that develop among the firms involved (Amaral and Uzzi 2007, Koza, and Lewin 1998, Wilkinson and Young 2002).

The problem of forming effective groups can be understood in terms of the theory of relational contracting in transaction cost economics (TCE). TCE focuses on matching governance structures to the types of transactions and interactions taking place (Williamson 1979, 1981, 1996). However, designing efficient governance structures is only part of the problem because it cannot be assumed that contractual provisions will actually be carried out due to post-contractual uncertainties and conflicts of interest, leading to potential problems of opportunism and shirking. As Oliver Williamson (1996, p 605) observes, "maladaptation in the contract execution interval is the principal source of inefficiency."

Implementation takes place over time through the interactions of the actors involved. This can be modeled as an iterated game in which the parties involved learn to communicate, coordinate and cooperate with each other in efficient manners. Experiments have shown that repeated interactions in a joint task result in the development of a task-specific form of communication that guides actions and responses in efficient ways (Weber and Camerer 2003, Selten and Warglien 2007). Further, relational contracting models show how actors can learn from their failures and develop efficient routines for coordinating actions (e.g. Chassang 2010, Blume and Franco 2007). An important factor underlying these results is the alignment of the interests of the parties involved, which reduces the risk of opportunistic behavior and shirking and promotes the development of more open communication, coordination and cooperation.

The degree to which alignment of interests develops depends on the nature of the coordination game and the incentives for each party. If each individual's outcomes are maximized when they behave in ways that maximize the group's total outcomes, there is an alignment of interests and it is only a question of learning what the right combination of actions is. If some parties gain more by acting in ways that damage group outcomes there is a misalignment of interests. A good example is the Prisoner's Dilemma game in which the greatest total gain comes from mutual cooperation whilst the greatest individual gain comes from defection against a cooperator. As a result there are mixed incentives to cooperate and defect. In a one shot game both parties defect to minimize their worst payoff and so lose the gains that would come from mutual cooperation. In repeated games cooperative strategies can prevail because cooperative behavior can be demonstrated and defection punished over time (e.g. Axelrod and Hamilton, 1981).

One way of aligning the interests of the parties is through implementing group based, as opposed to individual based incentives. Individual based incentive systems assess and reward individuals for their performance, whereas group based incentives assess and reward the group as a whole and divide the outcomes among group members in ways that are not directly related to their personal performance (see for example Weber and Camerer, 2003, and Selten and Warglien, 2007). Despite their demonstrated effectiveness, group based assessment and reward systems are not common in industry. In most firms individual based systems are the norm (Pfeffer 1998). The role of formal and informal group and team effects are largely ignored and unrewarded, despite the fact that much performance may be attributable to group interactions. As Henry Mintzberg observed:

> "We know that the most effective companies and organizations are those that embody the importance of being communities. ... But most conventional management practice and education has gone in completely the opposite direction. It's becoming more mercenary, more individualistic, less community oriented, and less nuanced." (cited in Kleiner 2010, p.2)

When group reward systems are used within firms to encourage more coordination and cooperation between people and departments, they are often criticized as they may potentially reward shirkers and free riders, as well as those who are actually driving the performance outcomes. As noted, however, group outcomes depend on interactions taking place over time in non obvious ways. Individuals who are apparent shirkers or free riders could potentially be playing essential roles in the group's overall functioning and performance. By not rewarding them or removing them from the group, group performance could be damaged. Building groups comprising only those who have individually performed well may produce dysfunctional, less cooperative, lower performing groups because the people making them up may not be prepared to contribute in ways that enhance total value.

A similar argument may be made about the manner in which firm performance is assessed and rewarded primarily in terms of individual firms' success rather than the performance of networks of interrelated firms in industries and value chains. This focus on the individual firm is reflected in theories that try to explain it only in terms of the behavior, resources, capabilities and orientations of individual firms and their managers and which exclude from explanations firms' relation and network contexts. Examples of such theories include: the resource based view (Barney 1991), resource-advantage theory (Hunt 1995), and the importance placed on various types of firm orientations, including market (Narver and Slater 1990), production (Nobel et al. 2002), technology and strategic (Gatignon and Xuereb 1997). Trade and industry policies also tend to focus on rewarding and boosting the performance of individual firms rather than interrelated networks and industries (e.g. Porter 1990, Wilkinson et al 2000).

The focus on individual-based incentive systems stems in part from a belief in the primary role of competition, in which only the fittest survive, in driving performance. This is an individualistic and adversarial approach in which business is seen as a form of war. But, as we have argued, individual and firm performance is to a large extent co-produced through people and firms interacting and working together, rather than by the independent actions of individuals and firms. Ensuring that people and firms work together effectively requires some alignment of interests which we will demonstrate can, in part, be achieved through group based assessment and reward systems.

## 2.1 The Evolution of Cooperation

While the emergence and value of cooperation among people is observed throughout the social realm, the reasons for this is unclear and remains a frequent topic of articles in leading scientific journals (e.g. Haidt 2007; Hruschka and Henrich 2006; Nowak 2006). There are three main explanations. Firstly, it is explained in terms of those involved having unique features that enable cooperators to recognize each other. This, however, is a not a stable solution, as it creates opportunities for non-cooperators to emerge who mimic, and so exploit, cooperators (Henrich 2004). The second type of theory is kinship based and explains cooperation in terms of the biological relation between two actors. People are more caring, self-sacrificing and cooperative towards others they are related to, such as their children, parents and siblings. The more closely related they are, i.e. the more genes they have in common, the more cooperative and self sacrificing they are (Axelrod and Hamilton 1981, Henrich 2004). Kinship based explanations, however, are not relevant for understanding cooperation in business, except in the relatively rare case of related individuals working in the same or different firms.

Third, the emergence of cooperation is explained as a consequence of the interactions taking place over time between individuals. Repeated interactions are, as already noted, different from one-off interactions. Cooperative strategies can emerge because of the shadow of the past (demonstration of past altruism) and the future (value of cooperation in subsequent interaction) (Axelrod 1984). This was demonstrated in computational simulations by Axelrod and Hamilton (1981), Lindgren 1997 and Hanaki et al. (2007). Computational work has also shown that cooperative strategies can evolve and develop in evolutionary settings that challenge the presumption that natural selection leads to the evolution of genes for selfish behavior (Bergstrom 2002).

## 2.2 Group Selection

A fourth way in which the emergence of cooperation may be explained has recently regained favor. Group selection, or more precisely multi-level group selection, focuses on the group as the unit of assessment, rather than the individual actor. This theory hypothesizes that groups, with some degree of continuity of membership and somewhat cooperative interaction, provide advantage to their members and outperform groups where survival is based on individual performance (Goodnight and Stevens 1997).

An illustration of the power of group selection mechanisms is provided by an experiment seeking the optimal method of egg production. Selective breeding of hens which lay the most eggs has been used for many years, resulting in birds that are aggressive with high mortality rates which undermine increased egg production. The seriousness of the problem is reflected in a popular Harvard Business Case, Optical Distortion Inc. (Clarke and Wise 2009). Research by Muir (1996) has showed that selecting the most productive groups of hens, rather than individuals, results in superior egg laying performance and normal life spans. He has since shown similar results in other types of animals and plants (Muir, 2005).

Group selection is not a new theory. It was initially embraced by evolutionary biologists (e.g. Wynne-Edwards 1962) then rejected (e.g. Maynard-Smith 1976, Williams 1966) due to the suspect methodological approaches of its adherents and because its explanations were less parsimonious than those of individual selection, A re-examination of the evolutionary equations underlying group selection, first proposed in the 1960s (Griffing 1967, Price 1970), led to a reappraisal of the theory and research. It was shown that experimental results demonstrating the superiority of individual selection did so due to the exclusion of interaction effects (Goodnight and Stevens 1997), in other words a random mixing of the population was assumed. Group selection theories, however, rely on this not being the case. Henrich (2004) has further developed the theory of group selection in terms of social and economic systems. Using a model based on the Price equations (1970), he explains the development of cooperative strategies in social systems in terms of within-group and between-group selection processes and establishes conditions for the persistence of altruism within a population even when egotism has strictly greater absolute fitness. He highlights a fundamental tension. Research in biological and social communities shows that, within groups, competitive behavior provides greater benefits to individuals than cooperative behavior but more cooperative groups outperform more competitive groups.

This line of work has important implications for the design of incentive systems to produce more cooperative intra and inter-firm relations. It suggests a mechanism by which more cooperative behavior and better group performance can be engineered through group as opposed to individual based incentive systems. The fundamental question is: Under what conditions do group based assessment, reward and selection systems do better than individual based systems in terms of the emergence of cooperative behavior and performance? The answer to this question has not yet been provided. Existing research has focused on the effects of individual selection mechanisms (e.g. Axelrod and Hamilton 1981). The effect of group selection has been largely overlooked with a few exceptions. Axelrod (1987) considered it in terms of spatial models whilst Bowles et al. (2003) show how the existence of certain types of group level structures may encourage the evolution of group beneficial traits. Work has also primarily focused on the IPD game rather than other types of games, which have different types of payoffs. No research has systematically compared the impact of group versus individual selection for different game conditions.

## 3 Modeling Group Interactions in terms of Iterated Games

In this section we present a simple model of the interaction of individuals under different social situations (games). We examine the effect of two different reward mechanism, group selection, in which the best performing groups are rewarded, and individual selection in which the best performing individuals are rewarded. We use an evolutionary learning algorithm to find the equilibrium strategies under each mechanism. Note, we do not argue that in reality individuals go through an evolutionary process, instead this is simply a mechanism to determine equilibrium behavior in different circumstances. As such this model is normative, rather than descriptive of the mechanism by which group selection occurs. By using this simple approach we are able to remove many incidental factors to show the circumstances under which each reward mechanism induces superior performance in both individuals and the team as a whole.

### 3.1 Model

We consider a model in which pairs of individuals within groups interact. The interactions take the form of two player, binary choice games. For ease of description we refer to the two actions as "cooperate" and "defect". They could, however, take any name, the important factor is the payoff for each pair of actions given in the payoff matrix and how this reflects the alignment (or not) of individual and combined outcomes.

The population consists of n players divided into m groups, each of equal size (n/m). In each generation every individual plays a specific 2 player game with every other individual within the group in turn. Each interaction last for r rounds. Here we implicitly model all members of the group as interacting equally frequently with all other members, however, this need not be the case. A network structure could be imposed on each group which would determine the individual interactions. It is known that the structure of relations within a group (e.g. Cummings and Cross, 2003) or units within an organization (e.g. Ethiraj and Levinthal, 2004) can influence its performance. Here we the simplest group structure in order to separate and demonstrate the effect of the reward and evaluation mechanism.

Each individual has a strategy dictating how they play the game. Strategies are represented as a string of zeros (defect) and ones (cooperate), which specify the players' response to all possible game situations. The set of situations a player may face is dependent on their memory length. With two possible actions, and a memory length of k, there are $2^k$ possible game situations the player must be able to respond to. In other words in each of k pervious time steps a players opponent will have played one of two different actions giving $2^k$ possible histories. For example, a strategy with memory length two must be able to respond to 4 different game histories of opponent actions i.e. DD, DC, CD and CC. Throughout this paper we use the contention that the most historically distant action is on the left with actions increasing in recentness to the right, for example DC means the opponent defected two periods ago, and cooperated in the last period.

This strategy, however, is not complete. Players actions depend on the history of their opponent's actions. At the start of the game, for the first k periods, the player does not have sufficient observations of their opponent's actions to be able to choose an action themselves. To resolve this a player's strategy also contains a fictitious history, k actions, which govern responses during the first k periods. For example with a memory of two, in addition to the 4 bits governing the players responses there are two additional bits giving the fictitious history (by convention we start the bit string with these values). A strategy of 010111 is composed of a fictitious history of 01 (first two positions) and 0111 (positions 3-6) corresponding to actions in the four states (described above). This means the individual defects if the other player defected in the previous two rounds (0 in position 3), cooperates after defection two periods ago and cooperation in the last period (1 in position 4), cooperates if their opponent cooperated two rounds ago but defected in the last (1 in position 5) and cooperates if the other player cooperated in both of the last two rounds (1 in position 6). A player's first move will be solely governed by their pre-assigned memory, in this case the memory is 01 which corresponds to D two periods ago and C one period ago. The agent therefore plays C, in the first period, the response to DC (fourth position) in their strategy. In the second period the second (more recent) value in the fictitious history (here the value 1) along with one round of real history are used. Let us assume the other player defects on the first round, the history will then be 10 (or CD) so the player will therefore play cooperate in the next round. After round two, decisions are solely based on the real history of play.

By representing the strategy as a single bit string we are able to use a simple Genetic Algorithm (Holland ,1975) to optimize behavior (in a similar manner to Axelrod, 1987, Midgley et al, 1997)[1]. At the start of the simulation individual strategies are randomly generated. In each generation each individual's score is calculated as the total value of the players payoffs from all games played against all other members of the individual's group. The group's score is the total score of all players within a group. Once every player has played every other player within their group a new generation of players is generated from the existing population. One of two alternative selection mechanisms is used to produce the groups. Under individual selection the best performing individuals are denoted as the n/2 individuals with the highest total scores in the population. These individuals are copied directly into the next generation. The remaining individuals are formed by randomly choosing pairs of the best performing individuals and creating a new strategy by combining the first part of one strategy with the second part of the other (crossover). In biology this mimics reproduction, whilst in social systems it may be thought of as a form of learning and sharing of strategies within a population. The break point for the parts is chosen rnaomdly from a uniform distribution. After crossover there is also a small probability (5%) of a single point mutation in a strategy in which a 0 flips to 1 or vice versa. The set of n new individuals are then randomly allocated to groups. Under group selection the m/2 best performing groups are selected and copied directly into the next generation. The remaining groups are formed by combining pairs of individuals selected from the best performing groups at random, irrespective of their individual performance.

---

[1] Miller (1996) uses an alternative automate mechanism, This approach allows more complex strategies at the expense of clarity of interpretation.

The genetic algorithm, by following this mechanism, selects the better individual strategies and allows them to spread throughout the population. Mating and mutation provide the mechanism by which successful strategies may be copied and new strategies developed. Over time this mechanism results in a population of strategies which are optimized to each other and the selection mechanism. The process of playing games and generating new populations is repeated 1000 times. Convergence occurs relatively early in this period (Figure 1) but continuing demonstrates the steady state. Games are played for r=200 rounds[2]. The population size, n=64, whist the number of groups, m=8. The memory length, k=3. Repeating the simulations with different numbers and sizes of groups, different population sizes and memory lengths did not qualitatively change the results but are available from the authors on request. All results are averaged over 100 repetitions with different random seeds.

Table 1 about here

We first describe the results for three specific iterated games prominent in the literature (payoff matrices are shown in Table 1). These were chosen to reflect three types of misalignment of interests between individual and combined outcomes, as well as being games that have frequently been used to model social and business interactions. In the second part of the paper we will consider a large space of games, the interactions and behaviors observed in the sample games will aid in understanding this larger space. The sample games are the Iterated Prisoners Dilemma (IPD) and two versions of the Chicken game (in some literatures this game is referred to as Hawk-Dove or Snowdrift). The prisoners dilema is a mixed motive game in which the Nash equilibrium for a one shot game is Defect-Defect (DD). Mutual cooperation results in the best combined outcome but the temptation to defect, to receive a higher individual payoff and avoid the lowest payoff, means it is not an equilibrium. The first Chicken game in Table 1, labeled

---

[2] Different numbers of rounds were examined but had no effect on the results.

Weak Chicken[3], produces the greatest combined payoff with combinations of Cooperate-Cooperate (CC) (4+4), Cooperate-Defect (CD) (1+7) and Defect-Cooperate (DC) (7+1). Under the second chicken game, Strong Chicken, the combined payoff for playing CD or DC is greater than that for always cooperating (e.g. 10+1 versus 4+4) or both defecting. Individual and combined outcomes are misaligned because one player receives the worst score under the outcome producing the highest combined return. In both of these games there are multiple Nash equilibriums, both the CD and DC outcomes provide no incentive for either player to change strategy. In a one shot game, however, there is no way for the players to determine which individual will take which role and so any payoff may occur. In repeated interactions other strategies may arise. The strong chicken game is an anti-coordination game, i.e. where complementary rather than the same actions are required to maximize combined payoffs, even though they result in greater variance in payoffs among the players. The weak chicken game differs in that the same combined payoff may be achieved by both players cooperating, even if a higher individual payoff may be achieved if one defected. Under repeated interactions equilibrium selection becomes an important aspect.

## 4 Results

Figure 1 shows the average payoff players receive from a single game at different times in the simulation for the genetic algorithm operating with group selection (left) and individual selection (right). Each figure shows the score averaged over 100 repetitions of the best (highest scoring) individual in the population, the worst (lowest scoring) individual and the average of all players for each of the three games.

We first consider the average payoffs of all individuals in the population. This measure captures the performance of the groups as a whole. As groups are the same size, a higher average payoff indicates a better group performance. For all games group selection produces better performance - higher average payoff – than individual selection.

---

[3] These games are similar to those labeled Weak and Strong alternation by Bednar et al (2010), however, the Weak Alteration and Weak Chicken differ importantly in their relative payoffs, we therefore do not adopt the same terms.

Figure 1 about here

It is perhaps not surprising that group selecting produces the highest average scores as a high average score translates into a high total group score. This selection mechanism explicitly selects those groups with the highest total score and so is effectively maximizing this value. Unexpected, however, in all cases, group selection also produces the best performing individuals. This is surprising as group selection does not attempt to maximize this aspect. In contrast, individual selection chooses the highest scoring individuals in each generation of the population and so would be expected to produce the best performing individuals overall. This underscores the importance of the group context when considering fitness and performance. Whilst individual selection chooses the best performing individuals it ignores the context, the group, in which they achieved the score. Group selection, however, is able to maintain strong environments which allow individuals to perform better. For example, within the strong chicken game individuals playing CD or DC achieve the highest combined score. In order to achieve this score one individual gets a high payoff whilst the other gets a low payoff (Table 1). Under individual selection only those obtaining the high payoff would survive, those obtaining the low payoff would be eliminated and so in subsequent generations payoffs would decrease as those with the same strategy playing against each other achieve a low payoff. Group selection is better able to support groups composed of varied individuals as evidenced by the wide spread between the best and worst individual scores under Strong Chicken. As a consequence, the best performing individuals are found under this mechanism with their performance being at the cost of others group members. The low scoring individuals could easily be regarded as underperformers but this is not correct because their behavior enables others to achieve greater gains for the group as a whole.

Whilst individual selection provides more impetus for high levels of individual performance it allows selfish, non-cooperative strategies to invade groups. This can reduce the absolute performance of the group and its best performing individuals. For example, a population in which all members of all groups cooperate in the IPD. If a single defector is added to the population, the group to which the individual is added will do worse than all other groups and be eliminated under group selection. In contrast, under individual selection, that defecting strategy will exhibit the best performance in the population and will be reproduced in the next generation.  Sophisticated strategies that discriminate between cooperators and defectors, playing the optimal response to each may limit the invasion of defectors, however, the random group formation means that a small percentage of defectors will achieve higher scores in their group and will persist under individual selection. A group that always cooperates will remain stable under group selection, achieving high fitness consistently.

Mean performance generally varies more under individual selection than group selection. This is because, under group selection, the environment in which strategies interact remains comparatively stable, whole groups are copied completely from generation to generation, allowing successful strategies to propagate and fixate in the population more quickly. In contrast for individual selection, a strategy that does well in one generation is not guaranteed to be so in the next generation due to the changing mix of strategies in its group. Consequently, evolution is noisier, exhibiting more variation.

### 4.1 Strategies

The results presented above demonstrate the ability of group selection to out-perform individual selection for three iterated games. In both the individual and group selection cases a particular strategy, such as tit-for-tat (TFT), does not come to dominate. Instead a variety of strategies continue to exist and evolve. The reason for this is that various combinations of strategies (the group's genotype) can produce the same observable behavior over time (the group's phenotype). For instance, in a group where there is consistently cooperative behavior, a strategy that would defect in response to a partner's defection, behaves in the same way as a habitual cooperator, though it possesses different response rules (genes) in some situations. In these circumstances it is difficult to distinguish between individuals who are always cooperative from those that give the appearance of being so. With continuous cooperation a strategy's responses to defection will be under no pressure to change. In successive generations they can continue to include in their strategies instructions to defect in response to defection without this affecting behavior or performance. In short, a bigger variety of genotypes (strategies) than phenotypes (behavior) can exist in a population because different mixes of strategies result in the same patterns of behavior. But if conditions change in a group, due to mutation for example, different outcomes and performance effects will occur because different strategies respond to the changed conditions in different ways. Some strategies lead behavior back to mutual cooperation, whereas others do not, resulting in a decline in performance and possible elimination.

In order to gain an appreciation of the range of strategies emerging and co-existing it is necessary to consider a large population rather than a single simulation result. The results are based on the set of strategies present in the 1000[th] generation of the simulations conducted above. Each cell gives the fraction of individuals who had a cooperate response in that location of their strategy. For clarity, the memory length three strategies are represented by strings of eleven zeros (D) or ones (C). The first three digits are the three period fictitious history whilst the next eight digits indicate how the strategy responds to each possible three period history i.e. DDD, DDC, DCD, DCC, CDD, CDC, CCD, CCC. For example, the strategy 111011111111 begins by cooperating (final digit=1), which is the response to CCC (first three digits=1), its initial memory, and it is not until it experiences three defections in a row that it defects (indicated by the zero in the fourth position). In all other situations it cooperates.

Table 2 about here

The strategies produced in the group selection and individual selection cases differ noticeably. Group strategies:

*Think Nicer:* Comparison of positions 1 to 3 shows that group selection strategies commence the game with a more positive attitude, being more likely to assume a history of.

*Act Nicer:* In general, the strategies evolved through individual selection are more likely to defect in response to a given pattern of behavior than group selection strategies. In particular, in group selection they are more likely to cooperate after three successive defections (55.5% versus 30.8%), which avoids getting caught in cycles of defection. They are more likely to continue to cooperate when the pattern is repeatedly cooperative (DCC, CCC) and are more forgiving, being more prepared to cooperate when there is cooperation after two previous defections (58% versus 26.4%).

*More Provocable*:  Group evolved strategies are not naïve cooperators, as they are just as likely as individual selection strategies to retaliate once the other starts to defect, i.e. responses to DCD and CDD. Group selected strategies have much in common with the characteristics of successful strategies identified by Axelrod's (1984) in an IPD setting:  nice, provocable, forgiving and clear.

### 4.2 Space of Games

So far we have compared group and individual selection for three games. There are, however, a vast number of games beyond those we have considered, each with associated incentives to cooperative or defect. In this section we examine a broad spectrum of games to determine the types in which individual or group selection are superior.

Rapoport and Guyer (1966) identify 726 strategically distinct 2x2 games in which individuals have weak preferences over outcomes[4]. For this analysis, however, the absolute values of the payoffs rather than simply their ordering is important in determining strategy success. This is because selection and so survival into the next generation depends on overall score across all interactions. Consequently there are an infinite number of possible games, making an exhaustive analysis impossible. In order to bound our analysis we restrict all payoffs to be integer multiples of 0.1 and to lie in the range [0,1]. This space consists of 14,641 games that show a great variety of different structures and include all of the standard orderings of payoffs. There are some types of games that cannot be represented in this format, for instance games that would produce lexicographic preferences (those in which the payoffs would be such that an individual would prefer to receive one particular payoff once rather than any of the others 200 times), nevertheless, this set is substantial and we believe sufficient for the analysis both in terms of its breadth and detail.

---

[4] If preferences are strong there are 78 distinct games.

In the following discussion we will refer to the payoffs in terms of the four possible outcomes for the row player, i.e. CC is "a", DC (where the row player plays D and the column player plays C) is "b", CD is "c" and DD is "d". Figure 2 demonstrates which selection mechanism produces the highest average performance and produces the best individual performance as a function of each payoff matrix in the space of games. In the figure, a=0.5 and b and c vary along the x and y axis respectively, whilst changes in d are represented in terms of different grids. For presentation we focus on a single value of "a", however, different values been analyzed, the results are qualitatively similar and are available on request.

Figure 2 about here

In Figure 2 a black cell corresponds to games where group selection scored higher, grey cells when individual selection scored higher and white cells depict no significant difference. There were 100 repetitions each for group and individual selection and all differences are significant with a probability of 0.01 based on a t-test. Figure 2A shows the results for average performance. There are four regions in which one selection mechanism does better than the other. The first region is the black triangular area approximately bounded by d<0.5, b>0.4, c<0.4, which includes games with payoffs structured like IPD (e.g. a=0.5, b=0.8, c=0.1 d=0.3) and Chicken (e.g. a=0.5, b=0.9, c=0.1 d=0.0). In this area the combined payoff for CC is the same or greater than CD/DC and DD has the lowest payoff. Consequently the reasoning discussed previously with regards to both of the Chicken and IPD games holds. Group selection allows the population to settle on an equilibrium that is beneficial to all. The payoff favors those that mutually cooperate and it is group selection that allows such strategies to survive in a group and be selected into future generations. The second region is the black triangular area approximately bounded by d>0.5, b<0.5, c>0.5, where the combined payoff for DD is the same or more than CD/DC and more than CC. This is equivalent to the first region except that the payoff matrix is mirrored, with DD being the equivalent of CC. Hence the explanation is the same.

The third region is a black triangular slice with its base at d=0 running between b=0.5, c=0.4 and b=1.0 c=0.6 and its point at d=0.4 b=0.9 and c=0.7. In some cases the pure CC equilibrium is optimal though in most cases the mixed CD is best. In both cases group selection is able to select the optimal mix of cooperators and defectors to maximize performance, whereas, under individual selection, this is not the case. Consider for example the game in which the payoffs are a=0.5, b=0.5, c=0.4, d=0.0. The best group payoff in this case is all cooperators but, if a defector is introduced into the population, they score as well as cooperators when playing cooperators (a=b) but when cooperators play them the cooperator gets the c score and b>c. This means the defector scores highest in the population, leading to its survival under individual selection, which reduces overall group performance.

The fourth region is the light grey triangular slice with its base at d=0 running between b=c=0.5 and b=c=1.0 and its point at d=0.9 b=c=1.0. Here the CD/DC option provides the greatest individual and combined payoff.

If the best combination is DC or CD there is a problem of establishing and maintaining a balance of cooperators and defectors in the group, in order to minimize CC and DD interactions. If the lesser payoffs for CC and DD are approximately equal, as is the case for this small region, the ideal balance between cooperators and defectors is 50:50. Individual selection achieves this self balancing more effectively than group selection. In order to understand why, consider a population of individuals who always cooperate or always defect. If the number of cooperators is greater than defectors a cooperator may increase their payoff, along with that of the group by changing to defection. This is because the number of CC interactions will decrease whilst the number of DC interaction will increase[5]. In the optimal 50:50 mix any player which changes strategy will decrease both their own and the group performance. Individual selection is able to quickly find this mix as the under-represented strategy will score more highly and so be more heavily selected in the next generation; the selection mechanism will favor a 50:50 mix. Group selection does not have this same equalizing force it forms new strategies from all those in the best groups, which will allow suboptimal strategy mixes to remain in the population for more periods.

---

[5] The number of DD interactions will also increase but by less than the decrease in CC interactions.

Figure 2B presents the results for the best performing individuals. Patterns here are less clear as there is considerably more variation in the best individuals in populations. Three areas, however, may still be observed. First, group selection produces the statistically best performing individuals in nearly half of all games. The results demonstrated earlier in the paper, showing that a good environment can benefit particular players, hold for a large proportion of the games. Broadly, there are two regions where the best individuals' performance occurs via group selection: for d<a the region b+c>1 and for d>a, c>d. In the first case the payoff for mixtures of strategies is higher than for pure CC or DD. Group selection is better able to maintain uneven group ratios between defectors and cooperators because, under individual selection, the better scoring strategy quickly drives the worse strategy out of the population. Group selection can produce individuals who score very highly within groups that are optimized to allow them to do so and which also maximize the group's overall score. A similar effect occurs in the second case where again there is a benefit from some individuals being cooperators in groups in which they are defected against.

Second, individual selection tends to produce the best individual in games where it also produced the highest average (as seen by the grey triangle b=c>0.5 d< 0.5), i.e. when individual and group interests are aligned. Third, there are other instances in which individual selection produces the best individual in a population though the areas in which this occurs are somewhat diffuse. The reason for this is that whilst a particular combination of strategies may result in a higher scoring individual, it is frequently the case that, as this individual spreads throughout the population, their performance is reduced. Consequently, the cases where individual selection produces the best individual are partially dependant on timing, i.e. at the moment of measurement there is a strategy or combination of strategies that makes one individual above averagely successful to the detriment of others.

## 5 Discussion

Group selection produces the best individual and higher average performance for a wider range of games than individual selection. The results show that group selection dominates in games where group and individual payoffs are not aligned, i.e. when one player's higher performance is at the expense of the group as a whole. Individual selection only does better when group and individual payoffs are aligned such that the combination of actions that maximize individual payoffs also maximizes group payoff. When interests are not aligned, group selection leads to the emergence and survival of cooperative ecologies of strategies in which some individuals do far better than others and the group as a whole benefits.  This is because group selection preserves group structure. Individual selection destroys such ecologies by weeding out strategies that are individually low performers but which benefit others. Under individual selection groups are formed of strategies which are individually high performing but which as a group may reduce both individual and group performance.

The different regions of results in Figure 2 suggest different types of group coordination situations found in business.  There is only one region in which individual selection produces superior group outcomes (shaded grey).  This is when individual and group interests are aligned such that the combination of actions that maximize individual payoffs also maximizes group payoff. This situation resembles simple or modular coordination tasks in which individuals specialize in different subtasks that can then be assembled or added together, as when groups of workers take turns in digging a hole, recording results, or serving different customers in a market (see the experiments of Selten and Warglien, 2007 and Weber and Camerer, 2003). Suboptimal results can still occur due to interaction effects, if members of the group attempt to help (or interfere) with another's work, if in doing so, they reduce their own performance more.

Group selection does better than individual selection when individual and group incentives are not aligned. To achieve the optimal performance some individuals must sacrifice their payoff for the good of the group as a whole[6]. A Pareto optimum results when it is possible to compensate those who sacrifice their own performance from the greater returns to the group as a whole. This resembles a problem in university departments, maximizing the number of quality publications whilst also completing teaching and administration. The department can benefit if some staff cooperate by doing more administrative duties, so long as others publish more high quality papers than those foregone by the "cooperators."

In general it is impossible to trace group outcomes to the separate contribution of individuals. A particular outcome may be highly dependent on one or more members sacrificing for the good of the group and it is important to distinguish between free riders and group sacrifices, as they are likely to resemble each other. In our simulations they are easier to distinguish because the group is modeled as a series of pair-wise interactions. For example we can observe when group outcomes improve or decline as a result of a change in an individual member's strategy. If group outcomes decline, there is indication of free rider type behavior dragging down group performance, but this performance depends on the mix of other strategies in the group. In real business situations such an analysis is more complex.

A number of other insights emerge from our simulations. First, they draw attention to the difference between the strategy mix in a group, i.e. its genotype, and the behavior resulting, i.e. its phenotype. Cooperative behavior can result even when the strategies involved include non-cooperative behavior rules because these rules are never activated in the group. Thus no one dominant strategy like TFT appears to emerge but the mix of strategies can still result in high performance. Rather than the genotype it is the phenotype – the behaviors that are actually observed which is important.

---

[6] This includes situations where CD/DC are the highest scoring but CC $\neq$ DD so an equal mix is not optimal.

Second, linked to the above point, our results show how the processes of strategic (or genetic) drift affect the evolution and survival of strategies. This is possible because there is an absence of selection pressure on particular response rules (or genes) in strategies. This is seen in Table 2, cooperation occurs under group selection in responses to DCD or CDD in approximately 50% of strategies. There is no pressure for always cooperating or defecting in response to these histories, the value in the strategy appears to be random. As a result non-cooperative strategies or naïve pure cooperators may arise due to random mutations, which in turn create opportunities for new types of strategies to emerge that counter or take advantage of them.

Third, the more evolutionary stable strategies emerging in the long run under group selection, have many of the characteristics first identified by Axelrod (1984) in relation to TFT (see Table 2 Section 4.1). In general, stable strategies are those that induce and sustain appropriate cooperative behavior in others, they are frequently: nice, provocable and forgiving.

Fourth, our results show the emergence of self-correcting mechanisms bringing groups back to better performing ecologies of strategies. This is seen under individual selection when individual and group interests are aligned such that CD and DC generate the greatest payoffs for individuals and the group. Here any deviation from an equal mix is quickly corrected by individual selection. A similar mechanism is observed in Tesfatsion (1997). Whether the existence of such patterns of change can be observed in business interactions is an interesting issue for future research.

## 6 Conclusion

In recent times management research and theory has focused attention on the role and importance of cooperation within and between firms as an important source of improved performance and competitive advantage. But there is scant research concerning the means by which cooperation within and among groups can be developed. Our results provide a systematic basis for understanding the conditions under which group based incentives systems outperform individually oriented systems in terms of the development of better performing groups and individuals and highlight the general conditions that bring about this superior performance. Group performance cannot be traced to simple additive contributions of each group member. Interactions matter. Performance is co-produced over time through interactions that in turn shape the composition and behavior of the group. When individual and group based interests are not aligned, individual evaluation and reward systems lead to the emergence of inferior group and individual performance, whereas under group based systems those involved learn to coordinate their actions and thereby boost overall performance. Forming and maintaining groups based on individual performance fails to take into account these interactions and leads to poorer performance compared to group based mechanisms unless individual and group interests are already closely aligned.

The capability of group assessment, reward and selection mechanisms to produce more cooperative, better performing groups is consistent with previous empirical results which examine the effects of different kinds of incentive systems in business. One way of aligning individual and group interests is through super-ordinate goals (e.g. Hunger and Stern 1976). These goals serve to make those involved reconceptualize the situation, reduce conflict and promote more cooperative behavior, which leads to better combined performance (e.g. Day, 1999). Katzenbach and Khan (2010) provide an example of how group based metrics and rewards were able to boost the performance of a failing company division.

These results have relevance for managers and policy makers in assessing and improving marketing and business organizations, relations and networks. Our results call into question the conventional wisdom and common practice in which group formation and reward structures are designed in ways that in the main favor selection of best performing individuals. This concern has been echoed by others and highlighted in recent developments in relational contract theory and research on the evolution of cooperation (e.g. Chassang 2010, Henrich 2004, Selten and Warglien 2007, Weber and Camerer 2003). This overemphasis on competition in part stems from the way that the effects of structural competition have been studied, i.e. situations in which two or more people vie for tangible or intangible rewards that are scarce.  Research has most often been conducted in laboratory settings that involve contrasts in performance between competitive and non-competitive individuals, rather than between competing individuals versus collaborating ones or between individuals and groups competing (Brown et al 1998).

There are concerns regarding the impact of free riders if they are not punished (e.g. Carpenter, et al 2004). While group rewards carry some risk of free riders and shirking, these must not be confused with another type of group member, the self-sacrificer, that plays a key role in enabling other members of a group to function much better. Group incentive systems allow such individuals to emerge and persist in groups.  A focus exclusively on individual incentives is likely to damage group functioning as this over-rewards the most visible performers, under rewards those who support them and fails to include in metrics the costs and damages associated with favoring the top performers (Sutton 2007). Apparently weak individual performance on some metrics may hide more complex interactions in which group performance depends in important ways on the behavior of those that do not appear to make a direct contribution to output. Removing or reprimanding such individuals may damage group functions in unintended ways. Forming new groups from the best individual performers in previous groups could well produce dysfunctional and/or poorly performing groups because previous high achievers are now required to act in a new group context in which more adversarial interactions and counterproductive outcomes arise.

For policy makers, our results reinforce the need to take inter-firm and industry relations and networks into account in developing trade and industry policy.  Our findings suggest the need to balance policies that reward individual firms with those that focus on the development of cooperative relations and networks that enable the group of firms to function more effectively (Porter 1990, Wilkinson et al 2000).

Finally our results illustrate the potential role of simulation models to examine the behavior of complex systems that are beyond the reach of traditional mathematical methods (Leombruni and Matteo 2005).

A number of fertile areas for future research present themselves. The model can be extended by examining the effects of alternative representations of interaction and the evolution of strategies. Additionally more complex group structures may be considered potentially incorporating networks of interactions and considering the interdependence between these and selection mechanisms.

This paper has considered group and individual selection as two alternatives, however, this does not have to be the case. Whilst group selection appears to outperform individual selection in many circumstances it may not fully differentiate in terms of rewarding self sacrificing behavior (good) as opposed to free riding behavior (bad). As such free riders may survive in a group over extended periods of time. One possible solution to this would be for firms to mix group rewards (encouraging cooperation) with individual incentives (encouraging effort). These two mechanisms interact in potentially complex ways so an interesting and very promising future research question is what is the optimal combination[7]. Our results are also useful in guiding future empirical research. Of particular importance is developing ways of identifying the self-sacrificers in groups, distinguishing them from free riders and studying the direct and indirect contributions they make to group functions in different contexts and how and why they play this role. Finally our results may inform the design of experiments and empirical studies to test the effect of incentive systems and the structure of relations in different interaction conditions.

---

[7] We thank an anonymous reviewer for this suggestion.

**Table 1:** Payoff matrices for three sample games (row player payoff, column player payoff)

|   | C | D |
|---|---|---|
| C | 4, 4 | 1, 5 |
| D | 5, 1 | 2, 2 |

a: Prisoners Dilemma Game

|   | C | D |
|---|---|---|
| C | 4, 4 | 1, 7 |
| D | 7, 1 | 0, 0 |

b: Weak Chicken Game

|   | C | D |
|---|---|---|
| C | 4, 4 | 1, 10 |
| D | 10, 1 | 0, 0 |

c: Strong Chicken Game

**Table 2:** Probability of a cooperating for strategies with memory length 3 in the final generation

| Position | Hist-3 | Hist-2 | Hist-1 | DDD | DDC | DCD | DCC | CDD | CDC | CCD | CCC |
|---|---|---|---|---|---|---|---|---|---|---|---|
| Group | 61.9% | 83.3% | 88.7% | 55.5% | 58.0% | 48.8% | 84.1% | 49.8% | 55.6% | 54.1% | 98.9% |
| Individual | 54.1% | 56.1% | 86.0% | 30.8% | 26.4% | 53.4% | 67.0% | 48.0% | 71.9% | 62.3% | 77.6% |

**Figure 1:** Scores of the best, worst and average individuals for the Prisoners Dilemma (top), Weak Chicken (middle) and Strong Chicken Game (bottom) under group (left) and individual selection (right).

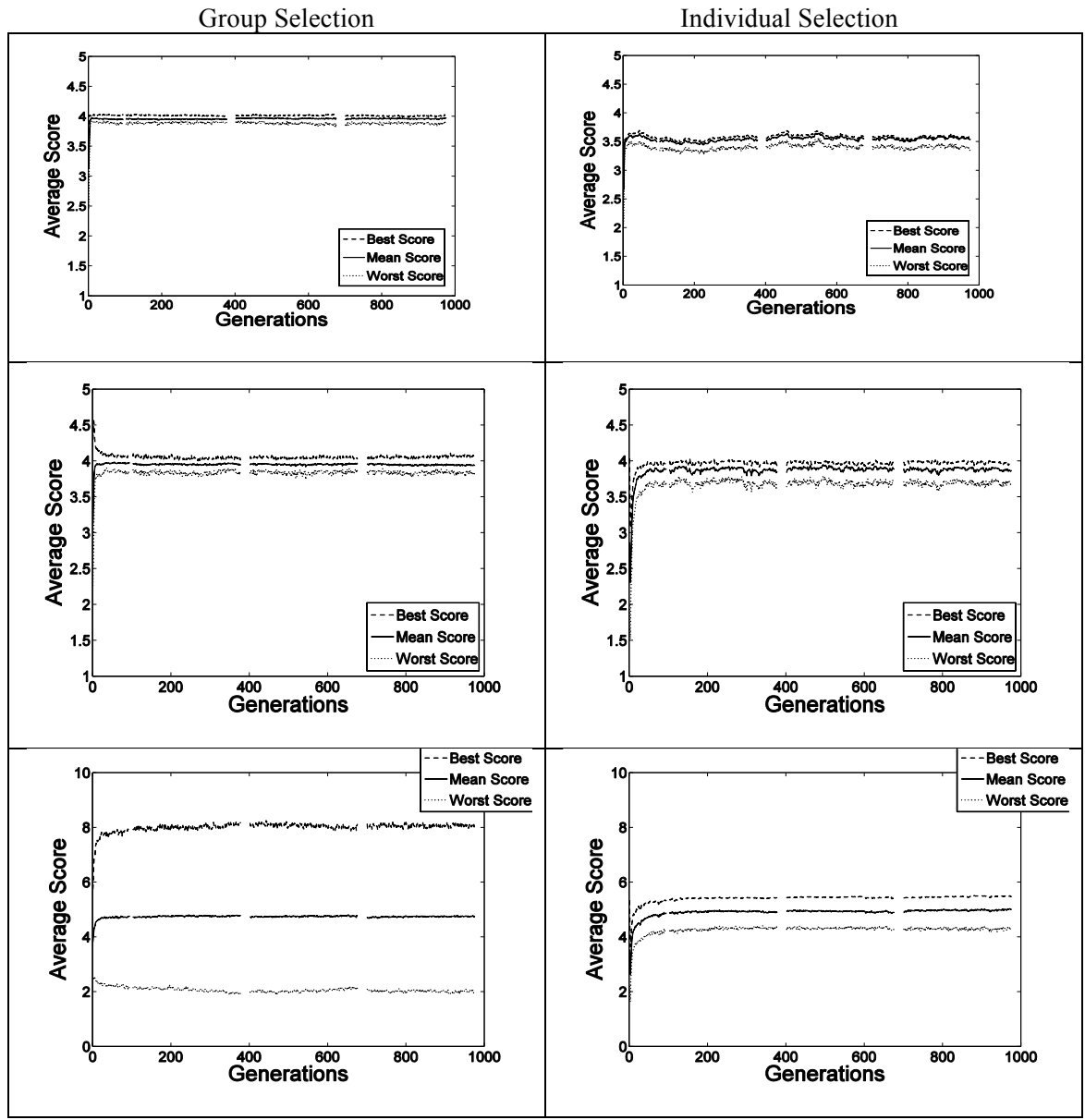

**Figure 2:** Comparison of group and individual selection mechanisms in terms of the final generation across a range of games. (payoff matrix: CC= a, DC=b, CD=c and DD= d. Simulations results with a=0.5, b on the x-axis, c on the y-axis, 100 repetitions each, p < 0.001 t-test). Top figure, mean performance of players. Bottom figure, performance of highest scoring players.

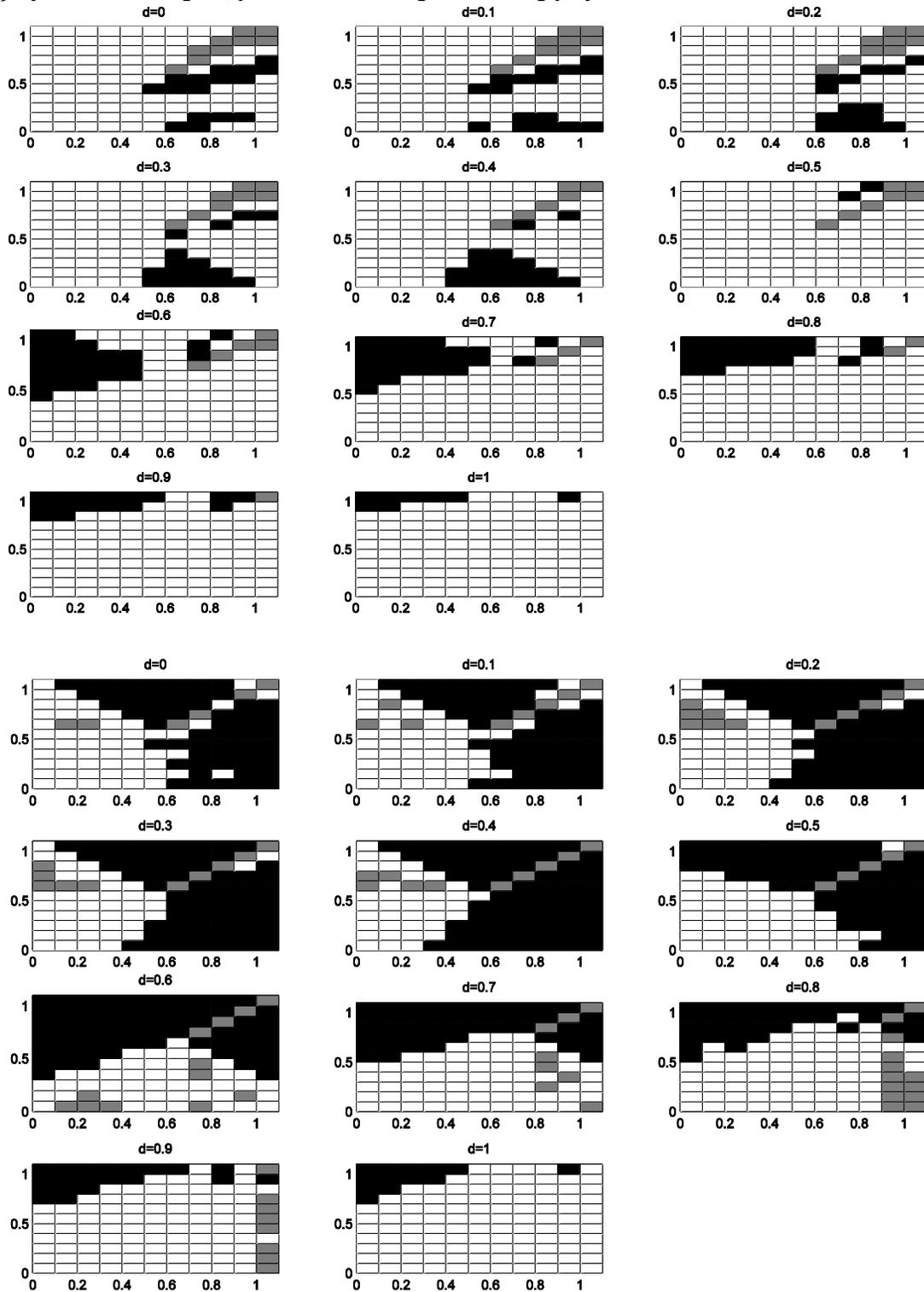